\documentclass[sigconf,screen]{acmart}
\usepackage{multirow} 
\usepackage{array}
\usepackage{graphicx}
\usepackage{tcolorbox}

\setcopyright{none}
\renewcommand\footnotetextcopyrightpermission[1]{}
\AtBeginDocument{%
  }

\AtBeginDocument{%
  }

\setcopyright{acmlicensed}
\copyrightyear{2018}
\acmYear{2018}
\acmDOI{XXXXXXX.XXXXXXX}
\acmConference[Conference acronym 'XX]{Make sure to enter the correct
  conference title from your rights confirmation email}{June 03--05,
  2018}{Woodstock, NY}
\acmISBN{978-1-4503-XXXX-X/2018/06}

\begin{document}

\title{StePO-Rec: Towards Personalized Outfit Styling Assistant via Knowledge-Guided Multi-Step Reasoning}


\author{Yuxi Bi}
\authornote{Equal contribution}
\affiliation{%
  \institution{College of Design and Innovation, Tongji University}
  \city{Shanghai}
  \country{China}
}
\email{yuxibi@gmail.com}

\author{Yunfan Gao}
\authornotemark[1]
\affiliation{%
  \institution{Shanghai Research Institute for Intelligent Autonomous Systems, Tongji University}
  \country{China}
}
\email{gaoyunfan1602@gmail.com}

\author{Haofen Wang}
\authornote{Corresponding Author}
\affiliation{%
  \institution{College of Design and Innovation, Tongji University}
    \country{China}
    }
\email{carter.whfcarter@gmail.com}

\renewcommand{\shortauthors}{Bi et al.}


\begin{abstract}
Advancements in Generative AI offers new opportunities for FashionAI, surpassing traditional recommendation systems that often lack transparency and struggle to integrate expert knowledge, leaving the potential for personalized fashion styling remain untapped. To address these challenges, we present PAFA (Principle-Aware Fashion), a multi-granular knowledge base that organizes professional styling expertise into three levels of metadata, domain principles, and semantic relationships. Using PAFA, we develop StePO-Rec, a knowledge-guided method for multi-step outfit recommendation. StePO-Rec provides structured suggestions using a scenario-dimension-attribute framework, employing recursive tree construction to align recommendations with both professional principles and individual preferences. A preference-trend re-ranking system further adapts to fashion trends while maintaining the consistency of the user's original style. Experiments on the widely used personalized outfit dataset IQON show a 28\% increase in Recall@1 and 32.8\% in MAP. Furthermore, case studies highlight improved explainability, traceability, result reliability, and the seamless integration of expertise and personalization.

\end{abstract}

\ccsdesc[500]{Information systems~Personalization}

\keywords{Personalized Outfit Styling Assistant, Fashion Knowledge Base, Explainability}


\maketitle



\section{Introduction}
\begin{figure}[htb]
    \centering
    \includegraphics[width=1\linewidth]{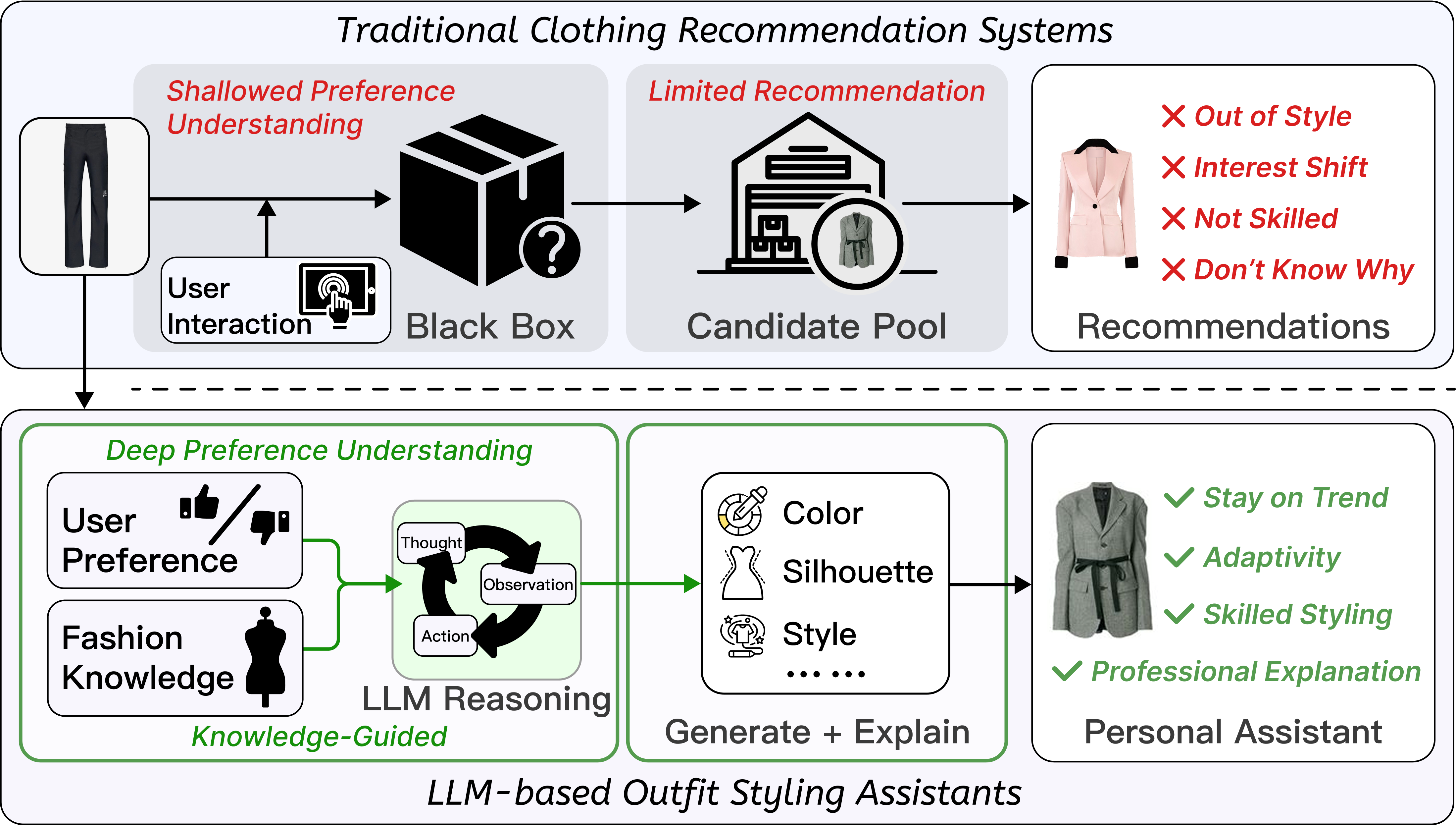}
    \caption{Comparison between traditional clothing recommendation systems and  LLM-based outfit styling assistants}
    \label{fig:intro}
\end{figure}
The application of artificial intelligence in fashion has revolutionized personalized fashion services~\cite{fashion_survey}. Advances in large language models (LLMs) and multi-modal technologies have upgraded the paradigm from simple product searches to intelligent, personalized outfit suggestions assistant. As shown in Figure~\ref{fig:intro}, traditional clothing recommendation systems face significant limitations, such as restricted candidate sets, weak user preference modeling, poor interpretability. This results in homogenized suggestions, cold-start challenges, and neglect of professional fashion knowledge~\cite{gong2023unified}. Black-box models further undermine trust due to their lack of interpretability~\cite{NOR}, leaving these systems poorly equipped to address rapidly evolving user expectations and fashion trends~\cite{pandit2020review}.

LLM-based methods, with advanced reasoning and deeper user preference understanding\cite{Chatrec,kalinin2024generative}, offer a promising alternative to traditional static recommendation systems. As personalized styling assistants, they can deliver tailored outfit suggestions, fashion advice, and real-time adaptation to changing trends and individual tastes. However, their limitations in applying domain-specific knowledge (e.g., color theory, body-shape coordination) and addressing rapidly evolving trends and cultural nuances highlight the need to overcome following key challenges: 

\textbf{(C1) Fine-grained Fashion Knowledge Modeling}. Current knowledge-based frameworks fail to integrate multidimensional information, such as garment attributes (e.g., color, silhouette), aesthetic principles (e.g., contrast balance), and scene style characteristics~\cite{ye2023show,zhou2021research}. This results in the long-established professional knowledge not being effectively utilized, undermining the credibility of recommendations. 

\textbf{(C2) Dynamic Preference-Knowledge Synergy Mechanism}. User decisions are shaped by long-term style preferences as well as short-term trend-following behaviors~\cite{ding2023personalized}, while domain knowledge evolves with industry. Crucially, achieving harmony between user preferences and outfit coordination knowledge is essential to prevent over-dependence on preferences that may lead to an information cocoon. 

\textbf{(C3) Explainable Recommendation}. Traditional feature importance methods fail to explain the implicit reasoning behind generative outputs. Although generative models bring expressiveness to recommendations, their opaque decision-making processes and potential hallucinations hinder user trust, often making it difficult to distinguish whether the LLM is providing evidence-based insights or confidently making things up.

To address the challenges identified, this paper proposes a unified framework. For knowledge modeling (C1), we construct PAFA (Principle-Aware Fashion), a comprehensive multi-modal knowledge base, which integrates hierarchical professional knowledge includes design principles, contextualized style matrices, and temporal trend fashion from  multi-sources. To address dynamic preference-knowledge coordination (C2), we design StePO-Rec, a tree-structured multi-step reasoning method that progressively refines outfit recommendations. It retrieves relevant professional knowledge from PAFA and combines it with user behavior data to adapt recommendations in harmony with individual preferences and evolving trends, further incorporating a Preference-Trend Re-ranking mechanism to amplify temporal awareness while maintaining personalization. To ensure both explainability and practicality (C3), we incorporated a decision path tracing system within StePO-Rec, which explicitly illustrates reasoning processes by linking each step to its theoretical and user-driven underpinnings.This bidirectional traceability promotes trust and transparency by offering users the dual perspectives of professional rigor and personal relevance.

Our main contributions can be summarized in threefold:

\begin{itemize}
    \item We introduce PAFA, a multi-modal fashion knowledge base that integrates professional theory, scene context, and temporal trends. By modeling knowledge hierarchically from metadata to principles to semantics, PAFA effectively supports multi-scale knowledge representation, enabling personalized fashion recommendations from macro-level style patterns to micro-level attribute details.
    \item We propose StePO-Rec, a tree-structured multi-step decision framework that combines preference understanding and knowledge constraints from PAFA. It enhances recommendation effect and adaptability by dynamically addressing dynamic user needs and fashion trends.
    \item Systematic experiments  demonstrate a 28\% improvement in Recall@1,and 11\% in Recall@10 for personalized recommendations. Case studies further illustrate the transparent decision-making process, highlighting the system’s interpretability and user-centric design.
\end{itemize}

\section{Realted Work}

\subsection{Personalized Outfit Recommendation}

Existing approaches to domain knowledge modeling using knowledge graphs primarily encode static attributes through expert-defined rules but often neglect multidimensional theories (e.g., aesthetic balance) and contextual factors (e.g., scenario-specific style matrices). 
For instance, Yang et al.~\cite{yang2019clothing} quantify outfit compatibility via expert-defined rules but focus on discrete attribute combinations, whereas Liu et al.~\cite{liu2012hi} employ dynamic graphs with user body parameters but overlook garment design principles and evolving trends, limiting cross-scenario adaptability. Collaborative filtering models like VBPR~\cite{VBPR} rely on historical user behavior to capture long-term preferences but struggle with shifting fashion trends and evolving interests. While GNN-based models such as GP-BPR~\cite{GP-BPR} and HFGN~\cite{HFGN} improve compatibility modeling and temporal shifts, their reliance on static graphs limits their ability to adapt to dynamic user preferences and trend signals. Interpretability in existing studies often leans on feature importance or templated explanations, failing to uncover implicit reasoning. Methods like NOR~\cite{NOR} generate explanations via cross-modal attention but lack alignment with recommendation logic, and BiHGH's~\cite{BiHGH} fixed rule libraries constrain transparency in recommendations.

\subsection{LLM-based Personalized Assistant}

LLMs have made notable strides in personalized knowledge modeling and retrieval, offering a strong foundation for addressing diverse user needs~\cite{zhang2024personalization}. By capturing user preferences and managing contextual information, recent research has emphasized user modeling~\cite{tan2023user,gpt4rec}, knowledge augmentation~\cite{IndustryScope,InstructRec}, and personalized retrieval~\cite{vemuri2023personalized} to improve understanding and responsiveness. USER-LLM~\cite{ning2024user} generates dynamic embeddings from users’ historical interactions, allowing real-time adjustments to outputs based on preferences. Similarly, analyzing users’ edits of model outputs to infer implicit preferences has shown promise\cite{gao2025aligning}. Retrieval-augmented generation (RAG)~\cite{gao2023retrieval} improves response accuracy by integrating external knowledge bases. Baek et al.~\cite{baek2024knowledge} proposed K-LaMP, a lightweight entity-centric personalization method that enhances LLMs by retrieving context-relevant entities from personal knowledge bases for personalized outputs. Similarly, LaMP~\cite{salemi2024lamp} evaluates and improves personalized output generation through diverse tasks and retrieval-augmented methods. In recommendation, combining user preferences with semantic retrieval has increased knowledge-matching precision~\cite{wang2022recindial}, while personalized recommendation explanations have strengthened user trust and experience~\cite{Chatrec}. Despite these advances, challenges remain. Current approaches focus on user modeling and history but struggle to incorporate domain-specific knowledge essential for complex tasks~\cite{zhao2024breaking}. Furthermore, generating explanations directly in natural language often results in inaccuracies and hallucinations, reducing their reliability in reflecting true decision-making processes~\cite{kalinin2024generative}.

\section{PAFA Knowledge Base}

\begin{figure*}
    \centering
    \includegraphics[width=1\linewidth]{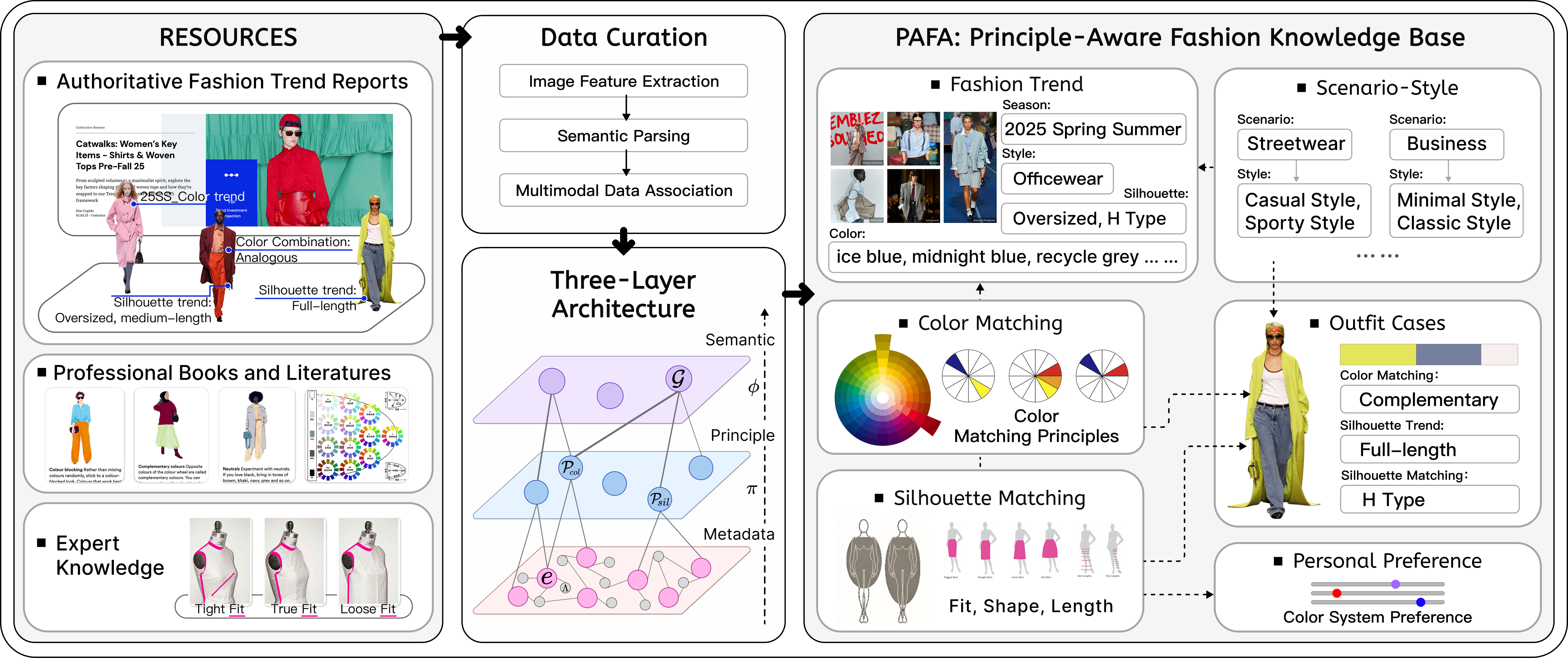}
    \caption{The PAFA Knowledge Base centralizes domain-specific knowledge by integrating fashion magazines, trend reports, styling guides, and designer insights through multi-step processing. It includes six sub-libraries: Fashion Trends, User Preference, Scene Styles, Color Coordination, Silhouette Matching, and Outfit Case Studies.}
    \label{fig:kb}
\end{figure*}

To address the challenges of multi-dimensional and cross-scale knowledge modeling in the field of outfits, we propose a Metadata-Principle-Semantic  three-layer architecture  knowledge base PAFA. As shown in Figure~\ref{fig:kb}, it achieves inter-layer coupling and cross-level knowledge integration from fine-grained attributes to high-level styles through a formalized definition $\mathcal{K} = (\mathcal{E}, \mathcal{A}, \mathcal{P}, \mathcal{S})$. The core components are described below.

\subsection{Metadata Layer}
As the data foundation of the knowledge base,  the metadata layer establishes fine-grained representations of clothing objects, which constructs the clothing feature space using the object entity set $\mathcal{E} = \mathcal{E}_o \cup \mathcal{E}_s$ and the metadata attribute set $\mathcal{A} = (\Lambda, \Theta, \Phi)$. Here, $\mathcal{E}_o = \{e_i\}_{i=1}^n $ includes atomic items (e.g., shirts, trousers) and composite coordination units (e.g., a three-piece suit), while $\mathcal{E}_s$ comprises semantic entities, such as styles (e.g., business, campus style).  

\emph{Metadata Attributes $\Lambda$}. Includes low-level attribute parameters, such as:  

\emph{Silhouette parameters} $\Lambda_{sil}$ (silhouette $v_{\text{sil}} \in \{H, X, A, O, Y\}$, fit $f_{\text{fit}} \in \{\text{tight}, \text{fit}, \text{loose}\}$).
\emph{Color attributes} $\Lambda_c$ (HCL coordinates $\langle H, C, L \rangle$, hue family, brightness, saturation).

\emph{Rule Attributes $\Theta$}. Define coordination constraints, such as the hue difference in a color wheel: $\theta_1: \Delta H(e_i, e_j) \geq 30^\circ$.

\emph{Semantic Attributes $\Phi$} Represent high-level style characteristics, such as $\phi_{\text{biz}} = \{\text{neutral color}, \text{straight cut}\}$.

A \emph{projection function}$\pi: \mathcal{E}_o \times \mathcal{A} \rightarrow \mathcal{E}_s$ maps attributes to semantic entities, enabling granular transformations and providing structured data to support higher-level reasoning.  

\subsection{Principle Layer}

The Principle Layer ($\mathcal{L}_{\text{principle}} = (\mathcal{P}, \mathcal{M})$) formalizes professional outfit design knowledge through the principle library $\mathcal{P} = \allowbreak \{\mathcal{P}_{sil}, \allowbreak \mathcal{P}_{col}, \allowbreak \mathcal{P}_{style}\}$ and the metric space $\mathcal{M}$. This structure ensures alignment across silhouette, color, and contextual dimensions for professional and adaptive design coordination.

\subsubsection{Principle Library ($\mathcal{P}$)}

The principle library includes three key aspects:
\emph{Silhouette Matching Principles ($\mathcal{P}_{sil}$)}. Quantifies fit, silhouette consistency, and length ratio in terms of overall morphological coordination.  
\emph{Color Matching Principles ($\mathcal{P}_{col}$)}. Ensures alignment of hue, brightness, saturation, and temperature for harmonious color schemes.  
\emph{Style Principles ($\mathcal{P}_{style}$)}. Defines scenario-specific constraints tailored to contextual needs. Each principle is represented by distinct sub-dimensions, and detailed information and examples are summarized in Table \ref{tab:principle-summary}.  

\subsubsection{Metric Space ($\mathcal{M}$)}

The metric space provides a quantitative evaluation framework:  
\emph{Style Compatibility ($\mu(s, e)$)}. Measures the proportion of satisfied principles under a target style, calculated as:  
\begin{equation}
  \mu(s, e) = \sum w_i \cdot I[\text{principle constraints } i],
\end{equation}
where $w_i$ are weights for principle $i$, and $I$ are indicator functions verifying compliance.  
\emph{Rule Satisfaction ($d(r, \mathcal{R})$)}. Tests for overall compliance by ensuring all principles for a specific rule set $\mathcal{R}$ are satisfied:  
\begin{equation}
d(r, \mathcal{R}) = \prod I[\text{principle } p \text{ is valid for set } \mathcal{R}].
\end{equation}

These metrics unify silhouette, color, and context principles under a cohesive, logical framework to optimize adaptability and design efficiency. 
\begin{table*}[htbp]  
\centering  
\caption{Summary of specific outfit styling principles in the principle layer}  
\label{tab:principle-summary}  
\renewcommand{\arraystretch}{1.1} 
\begin{tabular}{>{\centering\arraybackslash}p{2.2cm}|>{\centering\arraybackslash}p{2.3cm}|>{\centering\arraybackslash}p{2.5cm}|p{3.5cm}|p{4.6cm}}  
\toprule
\textbf{Principle} & \textbf{Definition} & \textbf{Sub-dimensions} & \centering\arraybackslash\textbf{Description} & \centering\arraybackslash\textbf{Example} \\ \hline  

\multirow{3}{*}{\textbf{Silhouette ($\mathcal{P}_{sil}$)}}  
& \multirow{4}{2.4cm}{$\mathcal{P}_{sil} = \sum_{i=1}^3 \rho_i F_i,$ \\ $F_i = \frac{1}{2}(S_i + P_i),$ \\ $\sum \rho_i = 1$}  
& Fit Degree ($F_1$)  
& The harmony of fit, avoiding proportional imbalance.  
& \multirow{3}{4.6cm}{A straight shirt with straight cropped trousers achieves proper fit ($F_1 = 0.85$), high shape consistency ($F_2 = 0.9$), and balanced length ratio ($F_3 = 0.8$). The overall score is $\mathcal{P}_{sil} = 0.85$, in compliance with H-Silhouette principle.}  
\\ \cline{3-4}  

&  
& Shape Type ($F_2$)  
& The harmony of shape types, avoiding conflicts or visual imbalance.  
& \\ \cline{3-4}  

&  
& Length Ratio ($F_3$)  
& The harmony of length, avoiding unflattering ratio.
& \\ \hline

\multirow{4}{*}{\textbf{Color ($\mathcal{P}_{col}$)}}  
& \multirow{3}{2.4cm}{$\mathcal{P}_{col} = \sum_{i=1}^4 \beta_i C_i,$ \\ $\sum \beta_i = 1$}  
& Hue Difference ($C_1$)  
& The angular distance ($\Delta H$) on the color wheel  
& \multirow{4}{4.6cm}{A light blue shirt with dark blue trousers exemplifies monochrome coordination principles with minimal hue difference $\Delta H = 8^\circ$, optimal brightness contrast $\Delta L = 15$, balanced saturation $S_{\text{ratio}} = 1.2$, and consistent temperature, archiving a overall score of $\mathcal{P}_{col} = 0.88$.} \\ \cline{3-4}  

&  
& Brightness Difference ($C_2$)  
& The measurable variation ($\Delta L$) in luminosity.
& \\ \cline{3-4}  
&  
& Saturation Ratio ($C_3$)  
& The comparative intensity ($S_{\text{ratio}}$) of color purity.
& \\ \cline{3-4}  

&  
& Temperature Consistency ($C_4$)  
& The uniform maintenance of color warmth or coolness  
& \\ \hline

\multirow{2}{*}{\textbf{Style ($\mathcal{P}_{style}$)}}  
& \multirow{2}{2.4cm}{$\mathcal{P}_{style}(s) = \{\theta_i\},$ \\ $S(s) = \prod \mathbb{I}[\theta_i \in s]$}  
& Scenario Compliance ($\theta_i$)  
& Binary evaluation indicator that assess whether all style requirements are met.  
& \multirow{2}{4.6cm}{Business contexts require formality rating $\geq 0.7$ and prohibit casual elements. An H-shaped suit with straight trousers in monochrome yields $S(s)=1$, while incorporating ripped jeans results in $S(s)=0$.} \\ \cline{3-4}  
&  
& Contextual Constraints ($\mathcal{P}_{style}(s)$)  
& Scenario-specific parameters expressed as logical conditions that define appropriate style boundaries.  
& \\ \bottomrule

\end{tabular}  
\footnotesize{  
\begin{flushleft}  
Note: $\rho_i$ and $\beta_i$ are weight parameters for silhouette and color principles, respectively; $S_i$ and $P_i$ are the independent scores of tops and bottoms in sub-dimension $i$; 
$\theta_i$ denotes specific constraint parameters (such as formality level, color restrictions, etc.), $S(s)$ is the scene compatibility score, and $\mathbb{I}[\theta_i \in s]$ is an indicator function that equals 1 if style requirement $\theta_i$ is present in style $s$, and 0 otherwise.  
\end{flushleft}  
}  
\end{table*}

\subsection{Semantic Layer}  

The semantic system $\mathcal{S} = (\mathcal{G}, \mathcal{C}, \Psi)$ provides an interpretable framework for style inference based on a bipartite style-rule graph. The core element of this framework is the style-rule graph $\mathcal{G} = (\mathcal{V}_g \cup \mathcal{V}_p, \mathcal{E}_{gp})$, where $\mathcal{V}_g$ represents style ontologies (e.g., business style, streetwear) and $\mathcal{V}_p \subseteq \mathcal{P}$ corresponds to rules within the principle layer (e.g., $\mathcal{P}_{\text{sil}}$, $\mathcal{P}_c$). The edges $\mathcal{E}_{gp}$ establish associations between styles and rules, with edge weights $w_{gp}$ computed as:  
\begin{equation}
w_{gp} = \lambda \cdot \frac{\sum_{c \in \mathcal{C}_s} I[p \in c.p \to]}{|\mathcal{C}_s|} + (1 - \lambda) \cdot \frac{\text{constraint strength of } p \text{ in } s}{\max(\text{constraint strength})},  
\end{equation}
where $\lambda$ balances the relative contribution of case frequency and constraint strength in defining the style-rule relationship. 

The semantic layer also introduces a case library $\mathcal{C} = \{c_k\}$, where each case $c_k = (\mathcal{I}_k, \mathcal{T}_k, \vec{s}_k, \vec{p}_k)$ consists of an image $\mathcal{I}_k$, a structured description $\mathcal{T}_k$, a style vector $\vec{s}_k \in [0, 1]^{|\mathcal{V}_g|}$ indicating the activation level of style ontologies, and a principle activation vector $\vec{p}_k$ encoding the adherence to specific rules. This case library serves as a knowledge base for contextualized style inference and rule compliance.  
The rule-style inference engine $\Psi$ functions as the generative mechanism for style determination given principle activations $\vec{p}$. Specifically, the style distribution is inferred as:  
\begin{equation}
P(s \mid \vec{p}) = \frac{\sum_j w_{sp_j} \cdot \vec{p}[j]}{\sum_{s'} \sum_j w_{s'p_j} \cdot \vec{p}[j]},  
\end{equation}
 
where $P(s \mid \vec{p})$ represents the likelihood of a style $s$ being activated given a specific principle vector $\vec{p}$, weighted by the corresponding style-rule relationships.  

Overall, the PAFA knowledge base includes 13 core apparel categories, including 4,800 standardized women's clothing entities with 16 attribute dimensions such as color, silhouette, design details, and style tags, covering 153 distinct structured attribute types. Additionally, it contains 477 unique outfit relationship types, annotated with multi-tags across color (5), silhouette (5), and style (28) dimensions. Trend content for 2025 Spring/Summer and Autumn/Winter includes 240 product trends, 120 scenario-style trends, 80 color trends, and 96 key silhouettes.

\section{Method}
\begin{figure*}
    \centering
    \includegraphics[width=1\linewidth]{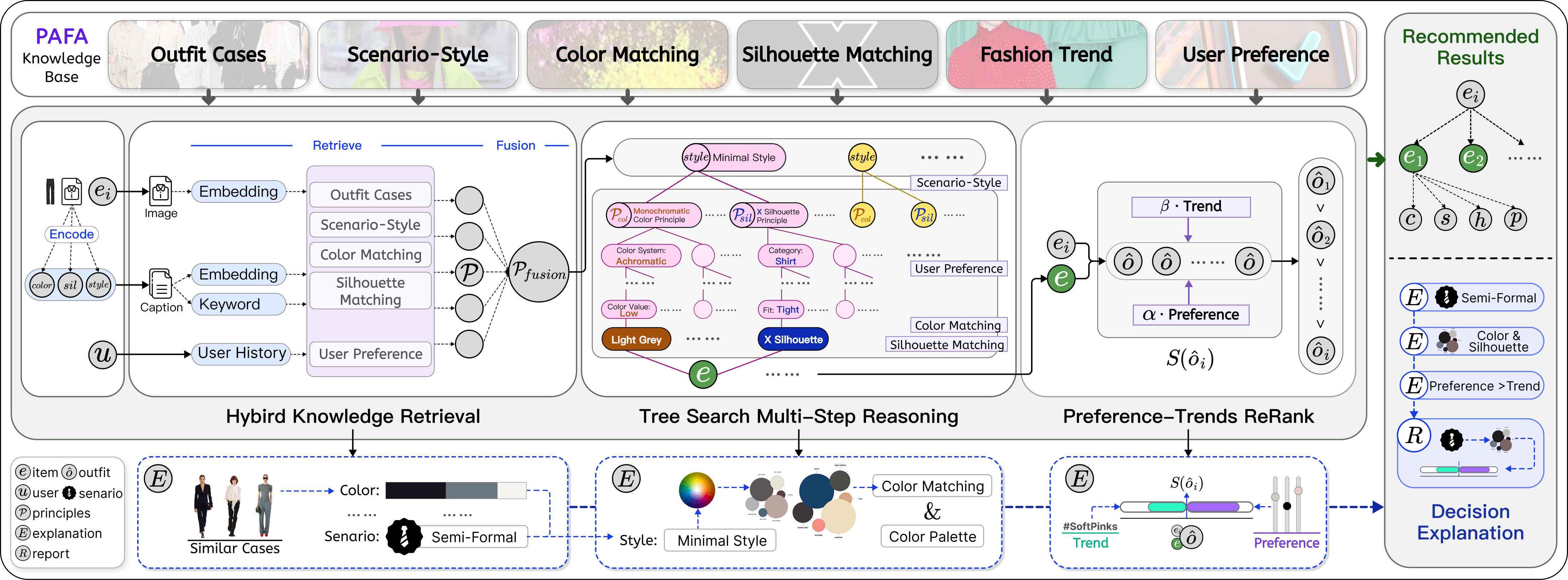}
    \caption{The StePO-Rec framework combines hybrid knowledge retrieval, tree-search multi-step reasoning, and preference-trends re-ranking. Using knowledge from the PAFA Knowledge Base, it deduces complementary garment attributes through tree-structured reasoning, with decisions at each node guided by dynamic context shaped by professional outfit styling rules and user preferences, ensuring interpretability and traceability at every step.}
    \label{fig:method}
\end{figure*}

\subsection{Problem Formulation}
We evaluate our personalized outfit assistant through the task of personalized outfit recommendation, which aims to recommend complementary clothing items (e.g., tops or bottoms) and their attributes based on fashion principles, trends, and user preferences.Formally, given a user set  $\mathcal{U}$, a collection of fashion items $ \mathcal{E}$, attributes $ \mathcal{A} $ with value space $\mathcal{V}$, and a principle knowledge base $\mathcal{K}$ (PAFA), the input is a quadruple $\mathbf{I} = (u, e_i, \mathcal{P}, H)$, where $u \in \mathcal{U}$ is the target user, $e_i \in \mathcal{E}$ is the anchor item, $\mathcal{P} \subseteq \mathcal{K} $ represents relevant fashion principles, and $ H = \{h_1, \dots, h_n\}$ reflects the user’s preferences. The goal is to generate an optimal attribute set $ V^* = \{v_j \mid a_j \in \mathcal{A}(e_o)\}$ for a complementary item $e_o \in E \setminus \{e_i\}$, meeting both fashion constraints and personalized needs.

\subsection{StePO-Rec Framework}
The Knowledge-Guided Multi-Step Reasoning Framework generates recommendations that balance fashion knowledge constraints with personalized adaptation. As shown in Figure \ref{fig:method}, it uses hybrid knowledge retrieval, multi-step reasoning, and reranking to provide professional, personalized and explainable recommendations.

\subsubsection{Hybrid Knowledge Retrieval}  
Using a multimodal hybrid retrieval method to extract both structured and unstructured knowledge from the PAFA knowledge base. On one hand, it helps quickly identify possible style scenarios through the PAFA knowledge system, while on the other hand, it provides relevant professional and personalized knowledge for subsequent multi-step reasoning.

Firstly, a VLM is used to  generates descriptions of the given garment image $I$, which are semantically parsed and transformed into structured queries to match the PAFA knowledge base format. The hybrid retrieval includes: \textbf{1) Typical outfits}. Top-K similar clothing items and corresponding outfits are retrieved from the outfit case database using cosine similarity between image embeddings $E_I$. Relevant outfit styling principles and their effects are identified through $
\mathcal{P}_v = \arg\max_{E_j \in \mathcal{C}} \left\{\frac{E_I \cdot E_j}{\|E_I\|\|E_j\|}\right\}_{j=1}^K
$. \textbf{2) Scene and style}. The scene and style inferred from the image are used to retrieve pairing rules from the scene and style library to ensure contextual compatibility. \textbf{3) Color and silhouette matching}. Relevant rules for color coordination and silhouette combinations are retrieved to ensure visually harmonious and cohesive outfit suggestions. \textbf{4) Personalized preferences}. The user’s historical outfit data is statistically analyzed to identify personal preferences, enabling more customized recommendations.  The final knowledge integration employs a LLM to reconciles multi evidence through $
\mathcal{P}_{fusion} = \Gamma_{LLM}(\mathcal{P}_v \cup \mathcal{P}_{style} \cup \mathcal{P}_{sil}\cup \mathcal{P}_{col} \cup \mathcal{P}_u)$ reducing redundancy and avoiding conflicts. This hybrid approach balances domain expertise with user adaptability, using structured retrieval to ensure contextual consistency and unstructured matching to capture nuanced visual-semantic relationships. 

\subsubsection{Tree Search Multi-Step Reasoning}
We present a principle-guided recursive tree-based approach that progressively refines the decision space from broad concepts to specific attributes. StePO-Rec models the intricate decision process as a hierarchical state-space exploration, with each state corresponding to a distinct styling decision point. The process is formally modeled as a directed tree $T = (V, E, \tau)$, where $V$ represents the set of decision nodes, $E$ denotes the directed edges between nodes, and $\tau: V \rightarrow \mathcal{T}$ maps each node to a decision type from the predefined type space $\mathcal{T} = \{\text{scenario}, \allowbreak \text{style}, \allowbreak \text{color}, \text{silhouette}, ...\}$. Each node $v_i \in V$ represents a specific state in the decision process, with the state space at node $v_i$ defined by its set of potential child nodes $\text{Child}(v_i) = \{v_j \in V | (v_i, v_j) \in E\}$.

The recursive state transition follows a Markov decision process that builds the solution tree incrementally. Starting from the root node $v_0$, the system recursively traverses down the tree, with each transition $(v_i, v_j) \in E$ representing a refinement of the recommendation space constrained by all prior decisions. This transition can be expressed as:

\begin{equation}
S_{j} = f(S_{i}, D_{i \rightarrow j}, \mathcal{K}_{\tau(v_j)}, \mathcal{P}(u))
\end{equation}

where $S_{j}$ is the state at node $v_j$, $D_{i \rightarrow j}$ represents the decision constraints imposed by the transition from $v_i$ to $v_j$, $\mathcal{K}_{\tau(v_j)}$ denotes the retrieved knowledge relevant to the decision type of node $v_j$, and $\mathcal{P}(u)$ captures the personalized preferences of user $u$.

The critical innovation lies in the dynamic context construction for each decision node. For any node $v_j$, it generate a context $C_j$ that encapsulates all relevant information required for making an optimal decision at that specific state:

\begin{equation}
C_j = \Phi([\Pi_{v_0 \rightarrow v_{j-1}}, \mathcal{K}_{\tau(v_j)}, \mathcal{P}_{\tau(v_j)}(u), \mathcal{V}(I)])
\end{equation}

where $\Pi_{v_0 \rightarrow v_{j-1}}$ represents the complete path of decisions from the root node to the parent of $v_j$, capturing all previously made decisions and their justifications. $\mathcal{P}_{\tau(v_j)}(u)$ extracts user preferences specifically relevant to the current decision type, and $\mathcal{V}(I)$ incorporates visual features extracted from input images $I$. The function $\Phi$ integrates these components into a coherent context  optimized for LLM reasoning.

The state transition probability at each node is then computed using an LLM-based decision function:
\begin{equation}
P(v_j|v_i, u, C_j) = \text{LLM}(C_j, A_{\tau(v_j)})
\end{equation}
where $A_{\tau(v_j)}$ represents the action space (possible choices) for the decision type at node $v_j$.

This dynamic context construction equips each node with position-specific knowledge in the decision tree. For instance, at a color principle selection node following choices of formal business and professional style, the context integrates formal dress codes, professional color guidelines, user color preferences, and reference image features—all assembled specifically for this decision state. The recursive tree construction continues through depth-first traversal until all required decision types have been addressed, yielding a complete path  $\Pi = \{v_0, v_1, ..., v_n\}$  that specifies all attributes of the recommended items. This approach ensures professionally sound yet personalized recommendations, with each decision honoring previous choices in the recursive constraint satisfaction process.

\subsubsection{Preference-Trend Collaborative Re-Ranking }
To enhance personalization, and trend sensitivity, we further introduce a Preference-Trend Collaborative Re-Ranking Mechanism after multi-step reasoning. The re-ranker builds upon the core concept of the ChatRec~\cite{Chatrec}, which utilize LLM to rerank the traditional recommender output based on user preference and external knowledge. Specifically, LLM constructs a comprehensive score \( S(\hat{o}_i) \) for each generated recommended result \( \hat{o}_i \) and performs re-ranking accordingly based on the user preference and trend information retrieved from PAFA. The scoring function can be formally expressed as:  
\begin{equation}
    \text{Score}(\hat{o}) = \alpha (u)\cdot \text{Preference}(\hat{o}) + \beta(u) \cdot \text{Trend}(\hat{o})
\end{equation}
Here, $\alpha$ and $\beta$ represent the weights for user preferences and trend relevance, which are dynamically adjusted based on user history. For example, users who prefer popular styles are assigned a higher $\beta$, while those valuing personal style consistency receive a higher $\alpha$. This module balances personal style stability with fashion trends, ensuring recommendations align with users' preferences while staying fashion-forward.

\subsubsection{Decision Explainability}
We construct an explainability module to generate user-oriented explanation reports that enhance recommendation credibility. Building upon the decision tree $T = (V, E, \tau)$ generated by StePO-Rec, the system traces and constructs decision paths $\Pi = \{v_0, v_1, ..., v_n\}$ within the "scenario-dimension-attribute" reasoning framework. For each node $v_i \in V$ along the path, we utilize customized prompt templates to extract decision content and corresponding justifications from each reasoning stage:

\begin{equation}
E_i = \Psi_{exp}(v_i, \mathcal{K}_{\tau(v_i)}, \mathcal{P}_{\tau(v_i)}(u), \Pi_{v_0 \rightarrow v_{i-1}})
\end{equation}

where $E_i$ represents the explanation content for node $v_i$. The complete explanation report $R$ is constructed by integrating all node explanations and relevant outfit combinations $\mathcal{C}$.

\begin{equation}R = \Omega(\{E_0, E_1, ..., E_n\}, \mathcal{C}, \mathcal{P}(u))
\end{equation}
where $\Omega$ is an integration function that organizes individual decision node explanations, relevant cases, and overall user preferences $\mathcal{P}(u)$ into a structured explanation report.

This module explicitly visualizes the recommendation logic and trade-offs between user preferences and professional constraints at each decision node. It supports bidirectional traceability, enabling users to either explore the recommendation path and how the system balances norms with individual traits, or reverse-query the theoretical basis and historical statistics behind key decisions. Through detailed explanations of each transition $(v_i, v_j) \in E$ in path $\Pi$, users can understand the rationale behind each decision step and its impact on the final recommendation outcome.

\section{Experiment}
To evaluate the proposed framework, we conducted a systematic series of experiments to address the following research questions:

\textbf{RQ1}: Can the proposed method effectively perform personalized outfit styling assist tasks?

\textbf{RQ2}: Does the integration of PAFA knowledge into StePO-Rec effectively enhance recommendation performance?

\textbf{RQ3}: Does StePO-Rec improve recommendation interpretability and enhance user trust in the results?

\subsection{Experimental Settings}
We preprocessed the widely-used  personalized outfit dataset IQON 3000~\cite{GP-BPR} by filtering for users with at least 250 outfits, removing incomplete outfits (single items or multiple bottoms). Data was split 80/20 for training and testing per user, preserving personalized preference modeling while preventing data leakage.For all retrieval-related tasks, we adopted a unified encoding approach, using the CLIP model for image encoding and the OpenAI Text-Embedding-3-small model for text encoding and retrieval.

The effectiveness of the proposed StePO-Rec framework was rigorously evaluated using two standard metrics in fashion recommendation: Recall@K (R@K) and Mean Average Precision (MAP). These metrics provide complementary insights into recommendation performance, with R@K emphasizing recall at specific ranking cutoffs and MAP capturing precision across the entire ranked list.   

\begin{equation}
\text{R@K} = \frac{1}{N} \sum_{i=1}^N \mathbb{I}_{\text{rel}}(i, \text{@K}), \quad   
\text{MAP} = \frac{1}{N} \sum_{i=1}^N \frac{1}{n_i} \sum_{k=1}^{L_i} \frac{\text{P}@k \cdot \mathbb{I}_{\text{rel}}(i,k)}{k}  
\end{equation}

Where, $N$ represents the total number of test samples used. For the $i$-th test sample: $n_i$ is the number of ground truth relevant items, $L_i$ is the length of the generated recommendation list, and the function $\mathbb{I}_{\text{rel}}$ is an indicator function with two forms: $\mathbb{I}_{\text{rel}}(i, \text{@K})$ equals 1 if any ground truth item appears in the top-$K$ recommendations and 0 otherwise; while $\mathbb{I}_{\text{rel}}(i,k)$ equals 1 if the item at rank $k$ in the recommendation list is relevant (i.e., matches the ground truth), and 0 otherwise.  

For R@K, results are reported for $K = \{1, 3, 5, 10\}$, reflecting the framework’s ability to surface ground truth items at critical early ranking positions. The MAP metric extends this evaluation by averaging precision scores at all relevant positions, where unretrieved ground truth items implicitly contribute zero precision. This dual assessment ensures comprehensive validation: R@K quantifies recall sensitivity under practical ranking constraints, while MAP penalizes positional inaccuracies in the recommendation list. 

\begin{table}[htbp]
    \centering
  \caption{Overall performance on personalized outfit recommendation task}
  \label{tab:main}
  \begin{tabular}{cccccc}
    \toprule
    Methods & R@1& R@3& R@5& R@10& MAP\\
    \midrule
    GLM-4-plus & 0.07 & 0.34 & 0.38 & 0.46 & 0.39\\
    ReAct (GLM-4-plus) & 0.06 & 0.39 & 0.47 & 0.53 & 0.39\\
    GPT-4o-mini & 0.06 & 0.23 & 0.28 & 0.49 & 0.33\\
    ReAct (GPT-4o-mini) & 0.07 & 0.22 & 0.27 & 0.36 & 0.32\\
    GLM-4v-plus & 0.06 & 0.20 & 0.27 & 0.28 & 0.41\\
    Qwen-vl-max & 0.11 & 0.52 & 0.60 & 0.60 & 0.46\\
    DeepSeek-R1 & 0.14& 0.32 & 0.38 & 0.56 & 0.33\\
    Qwen-QWQ& 0.26& 0.31& 0.54& 0.54& 0.48\\

    \midrule
    NaiveRAG & 0.37& 0.57& 0.61& 0.61& 0.55\\
    Re-ranking  & 0.43& 0.60& 0.63& 0.63& 0.64\\
    GraphRAG  & 0.31& 0.58& 0.64& 0.64& 0.41\\
    NaiveRAG w/o Pref& 0.29& 0.43& 0.58& 0.58& 0.46\\
    Re-ranking w/o Pref& 0.33& 0.33& 0.43& 0.57& 0.34\\
    GraphRAG w/o Pref& 0.14& 0.42& 0.59& 0.59& 0.35\\
    \midrule
    StePO-Rec (ours) & \textbf{0.55} & \textbf{0.69} & \textbf{0.71} & \textbf{0.71} & \textbf{0.85}\\
    \bottomrule
  \end{tabular}
\end{table}

\subsubsection{Baselines}  

To comprehensively evaluate the performance of \textit{StePO-Rec}, we compared it against baselines from six representative categories: 1) \textbf{LLMs} (e.g., GLM-4-plus, GPT-4o-mini): Evaluates the performance of general-purpose language models under non-specialized conditions. 2) \textbf{VLMs} (e.g., GLM-4v-plus, Qwen-vl-max): Assesses multi-modal semantic understanding by leveraging structured data and raw visual images for fashion recommendations. 3) \textbf{LRMs} (e.g., DeepSeek-R1): Tests the intrinsic logical reasoning capabilities of large reasoning models when provided with fundamental outfit principles. 4) \textbf{RAG Methods}.By retrieving knowledge from PAFA, LLMs are provided with specialized fashion coordination knowledge. Following Gao et al.~\cite{gao2023retrieval}, we use the NaiveRAG m a two-stage RAG method~\cite{mortaheb2025re} with re-ranking for comparison, as well as the KG-based GraphRAG~\cite{graphrag}.5) \textbf{Agent Frameworks}. ReAct~\cite{react}combines reasoning and action in an iterative process, allowing agents to generate thoughts, retrieve relevant information, and take appropriate actions. 

\subsection{Results and Analysis}

\subsubsection{Performance Comparsion (RQ1)}
StePO-Rec was evaluated using the IQON-3000 dataset (Table ~\ref{tab:main}) and demonstrated significant improvements across key metrics, achieving a Recall@1 of 0.55 and a MAP of 0.85—28.3\% and 32.8\% higher than the best-performing baseline, Two-Stage RAG with reranking. Its Recall@1 was also 12\% higher than NaiveRAG in the top-1 recommendation.The high MAP score (0.85 vs. NaiveRAG’s 0.64) further emphasizes StePO-Rec’s precision across the entire recommendation list, showcasing its robustness in aligning long-tail user preferences with evolving fashion trends through the preference-trend reranking mechanism.

General-purpose LLMs (e.g., ChatGPT, GLM) struggled due to a lack of domain-specific fashion knowledge, yielding poor Recall@1 results below 0.10. In contrast, StePO-Rec, leveraging professional fashion knowledge via its PAFA module,  proving the importance of domain expertise. Multimodal models like GLM-4v-plus and Qwen-vl-max, while capable of processing images and text, lack structured fashion knowledge and combinatorial reasoning, limiting their ability to balance "fit" and "profession".

RAG methods (e.g., NaiveRAG, GraphRAG) partially captured elements like scenarios, silhouettes, and color combinations outlined by PAFA, improving performance (NaiveRAG: Recall@1 = 0.37). However, their reliance on single-turn or fixed graph-based retrieval restricts iterative, adaptive reasoning and limits personalization. Agent-based frameworks like ReAct offered only slight improvements over standard large language models (e.g., Recall@1 increased from 0.06 to 0.07), showing that excessive flexibility can compromise precision. This underscores the difficulty of aligning freeform reasoning with domain-specific constraints.

\begin{table} [htbp]
    \centering
  \caption{Ablation Study}  
  \label{tab:ablation}  
  \resizebox{0.5\textwidth}{!}{  
    \begin{tabular}{cccccc}  
      \toprule  
      Baseline & R@1& R@3& R@5& R@10& MAP\\
      \midrule  
      w/ only Reasoning & 0.19 & 0.28 & 0.51 & 0.51 & 0.46\\
      w/o Knowledge Rertrieval & 0.34 & 0.47 & 0.49 & 0.49 & 0.36\\
      w/o Re-Ranking & 0.45 & 0.67 & 0.70 & 0.71 & 0.57\\
      StePO-Rec & 0.55 & 0.69 & 0.71 & 0.71 & 0.85\\
      \bottomrule  
    \end{tabular}  
  }  
\end{table}  

\subsubsection{Ablation Study (RQ2)}
To evaluate the contribution of each module, we conducted ablation experiments (Table~\ref{tab:ablation}). Removing hybrid knowledge retrieval significantly reduced performance. Without domain-specific fashion knowledge from PAFA, the model's ability to infer styles, adapt to color schemes, and align with user preferences weakened, resulting in a 38.2\% drop in R@1 (0.55 → 0.34). This demonstrates that professional knowledge is critical for guiding decision tree reasoning. Removing the re-ranking module caused a 32.9\% decrease in MAP (0.85 → 0.57), as the model struggled to adapt to user preferences and trends despite maintaining basic outfit compatibility. This highlights the module's importance in balancing personalization and trend sensitivity. Finally, relying solely on reasoning or retrieval led to drastic performance declines (e.g., R@1 = 0.19 for reasoning alone), showing that integrating retrieval-based knowledge with reasoning achieves the best balance between expertise and adaptability. In summary, these experiments confirm that combining hierarchical knowledge retrieval, recursive reasoning, and dynamic re-ranking is essential to bridge the gap between theoretical fashion principles and user needs.

\begin{figure*}[htbp]
    \centering
    \includegraphics[width=1\linewidth]{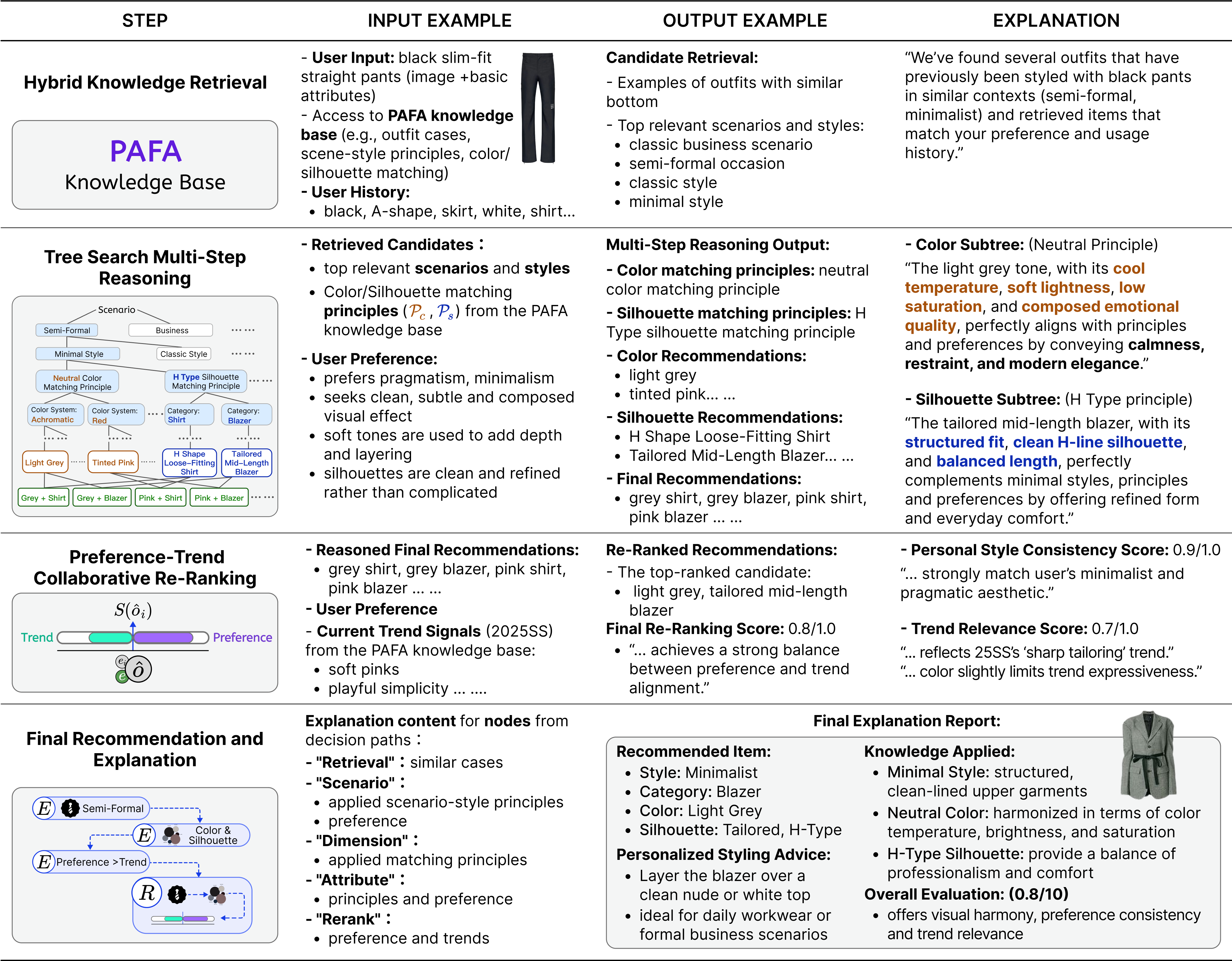}
    \caption{Case study of StePO-Rec}
    \label{fig:case}
\end{figure*}

\subsection{Case Study (RQ3)}  

In this case study (Figure~\ref{fig:case}), we demonstrate how StePO-Rec applies the PAFA knowledge base to generate a personalized styling recommendation, using \emph{black slim-fit straight pants} as the anchor item. While the PAFA knowledge base itself is built "from the bottom up"---\emph{metadata} to \emph{principles} to \emph{semantics}---the actual recommendation process proceeds "from the top down".

\emph{Step 1: Identify Style Scenario (Semantic Layer).}  
StePO-Rec begins by retrieving relevant cases, which includes annotated images, styles (e.g., business, streetwear), and principle activation indicating applied principles (e.g., silhouette, color guidelines). By matching the user’s past preferences (e.g., subdued color palettes) to similar cases in the library, the system infers a likely \emph{semi-formal} or \emph{business-casual} scenario for the anchor pants. 

\emph{Step 2: Apply Professional Principles (Principle Layer).}  
After identifying the styles, StePO-Rec uses the \emph{principle layer} to apply  guidelines for silhouette and color matching. Structured or slightly loose-fitting tops are recommended to balance slim-fit straight pants, while neutral or soft accent colors like light grey or muted pink ensure business-appropriate subtlety and reflect seasonal trends. This approach grounds recommendations in professional, domain-specific principles rather than purely data-driven methods.

\emph{Step 3: Refine Concrete Attributes (Metadata Layer).} The system utilizes the metadata layer to define item-level details, such as specifying a "light grey blazer with H-type tailoring" or a "lightly loose-fitting blouse" to pair with the user’s black slim-fit pants.This layer finalizes the "top-down" process, linking semantic style inference, principle-driven rules, and detailed attribute refinement.

\emph{Preference-Trend Collaborative Re-Ranking.}  
After assembling a set of potential outfits, StePO-Rec applies a re-ranking that integrates both personal style signals (e.g., minimalism) and industry trends (e.g., Spring-Summer updates). Outfits aligned with the user’s subtle palette preference and the recommended semi-formal ambiance receive higher final rankings.

Through this layered procedure, StePO-Rec incorporates a \textbf{Semantic Layer} for scenario-level style inference based on a bipartite style-rule graph and case library, a \textbf{Principle Layer} for rigorous professional guidelines on silhouette and color schemes, and a \textbf{Metadata Layer} for final item attribute selection.

\section{Conclusion and Future Work}
This paper introduces StePO-Rec, a personalized outfit styling assistant guided by domain-specific knowledge from the PAFA knowledge base, which uniquely integrates professional styling principles, user preferences, and fashion trends. By employing a tree-structured multi-step reasoning framework, the system dynamically constructs recommendations that are both traceable and explainable, balancing professional knowledge with personal preferences. While effectively bridging personalized outfit styling with fashion expertise, several areas invite further exploration. To advance, further efforts will focus on scaling PAFA to handle diverse datasets and dynamic trends, integrating human-in-the-loop mechanisms for feedback from both users and experts, and enhancing the efficiency of recursive reasoning and collaborative re-ranking for large-scale deployment.

\bibliographystyle{ACM-Reference-Format}
\bibliography{stepo_rec}
\appendix
\newpage

\section{Symbol and Notation}

The relevant symbols and notation used in this article are summarized in the Table~\ref{tab:symbol-ref}.

\begin{table*}[htbp]  
\centering  
\caption{Symbol Reference Table}  
\label{tab:symbol-ref}  
\resizebox{\textwidth}{!}{
\renewcommand{\arraystretch}{0.8}  
\begin{tabular}{p{3.5cm} p{3cm} p{7.5cm} p{3cm}}  
\toprule  
\textbf{Symbol} & \textbf{Type} & \textbf{Definition / Meaning} & \textbf{Occurrence} \\
\midrule  
\multicolumn{4}{l}{\textbf{Global Knowledge Base}} \\
\(\mathcal{K}\) & Knowledge Base & Overall structure: \(\mathcal{K} = (\mathcal{E}, \mathcal{A}, \mathcal{P}, \mathcal{S})\). & Chapter 3, Figure~2 \\
\midrule  
\multicolumn{4}{l}{\textbf{Metadata Layer}} \\
\(\mathcal{E}\) & Entity Set & Complete entity set: \(\mathcal{E} = \mathcal{E}_{o} \cup \mathcal{E}_s\).   
\(\mathcal{E}_{o}\) denotes objects (shirts, pants, etc.) and composite units, while \(\mathcal{E}_s\) denotes semantic entities (styles, scenarios). & Metadata Layer \\
\(\mathcal{E}_o\) & Object Entities & Atomic items (e.g., shirts, trousers) and composite outfit units (e.g., a three-piece suit). & Metadata Layer \\
\(\mathcal{E}_s\) & Semantic Entities & High-level semantic entities (e.g., “business style,” “campus style”). & Metadata Layer \\
\(\mathcal{A}\) & Attribute Set & Attributes in the metadata layer: \(\mathcal{A} = (\Lambda, \Theta, \Phi)\). & Metadata Layer \\
\(\Lambda\) & Low-Level Attributes & Includes silhouette parameters (\(\Lambda_{\mathrm{sil}}\)) and color attributes (\(\Lambda_{c}\)). & Metadata Layer \\
\(\Theta\) & Rule Attributes & Constraints for garment coordination (e.g., \(\Delta H(e_i, e_j) \geq 30^\circ\)). & Metadata Layer \\
\(\Phi\) & Semantic Attributes & High-level style characteristics, e.g., \(\phi_{\mathrm{biz}}=\{\mathrm{neutral\ color},\; \mathrm{straight\ cut}\}\). & Metadata Layer \\
\(\pi: \mathcal{E}_o \times \mathcal{A} \to \mathcal{E}_s\) & Mapping Function & Projects object entities and their attributes into semantic entities, supporting higher-level reasoning. & Metadata Layer \\
\midrule  
\multicolumn{4}{l}{\textbf{Principle Layer}} \\
\(\mathcal{P}\) & Principle Library & Professional outfit design principles: \(\{\mathcal{P}_{\mathrm{sil}},\; \mathcal{P}_{\mathrm{col}},\; \mathcal{P}_{\mathrm{style}}\}\). & Principle Layer \\
\(\mathcal{P}_{\mathrm{sil}},\; \mathcal{P}_{\mathrm{col}},\; \mathcal{P}_{\mathrm{style}}\) & Principle Sub-libraries & Sub-libraries for silhouette, color, and style constraints, respectively. & Principle Library \\
\(\mathcal{M}\) & Metric Space & Quantitative framework (e.g., silhouette scores, color harmonies, style compliance). & Metric Space \\
\(\mu(s, e)\) & Style Compatibility & Measures how well an item \(e\) aligns with style \(s\) according to satisfied principles. & Metric Space \\
\(d(r, \mathcal{R})\) & Rule Satisfaction & Checks if rule \(r\) complies with all relevant principles in \(\mathcal{R}\). & Metric Space \\
\midrule  
\multicolumn{4}{l}{\textbf{Semantic Layer}} \\
\(\mathcal{S}\) & Semantic System & \(\mathcal{S}=(\mathcal{G}, \mathcal{C}, \Psi)\), providing an interpretable framework based on a style-rule bipartite graph. & Semantic Layer \\
\(\mathcal{G} = (\mathcal{V}_g \cup \mathcal{V}_p,\; \mathcal{E}_{gp})\) & Style-Rule Graph & Bipartite association among style ontologies \((\mathcal{V}_g)\) and principles \((\mathcal{V}_p \subseteq \mathcal{P})\). & Semantic Layer \\
\(w_{gp}\) & Edge Weight & Reflects both case frequency and constraint strength in style-rule links. & Semantic Layer \\
\(\mathcal{C}\) & Case Library & A set of style cases \(\{c_k\}\), each with image \(\mathcal{I}_k\), textual description \(\mathcal{T}_k\), style vector \(\vec{s}_k\), and principle vector \(\vec{p}_k\). & Semantic Layer \\
\(\Psi\) & Inference Engine & Produces style distributions \(P(s \mid \vec{p})\) based on principle activations. & Semantic Layer \\
\midrule  
\multicolumn{4}{l}{\textbf{Method}} \\
\(T = (V, E, \tau)\) & Decision Tree & The tree structure for multi-step reasoning: \(V\) = node set, \(E\) = edges, and \(\tau\) is node type (scene, color, etc.). & Tree Search (Method) \\
\(\Pi = \{v_0,\ldots,v_n\}\) & Decision Path & A complete path (root \(\to\) leaf) that captures all decisions (e.g., scenario \(\to\) style \(\to\) color). & Tree Search (Method) \\
\(C_j\) & Dynamic Context & Context at node \(v_j\), integrating prior steps \(\Pi_{v_0 \rightarrow v_{j-1}}\), relevant knowledge, and user preferences. & Tree Search (Method) \\
\(\mathcal{P}(u)\) & User Preferences & Personalized preferences for user \(u\), guiding specific decision nodes. & Tree Search (Method) \\
\(P(v_j \mid v_i, u, C_j)\) & Transition Probability & Probability of moving from node \(v_i\) to \(v_j\), given user \(u\) and context \(C_j\). & Tree Search (Method) \\
\(\Gamma_{\text{LLM}}\) & Knowledge Fusion & LLM-based function that merges principle sets from multiple sources. & Hybrid Knowledge Retrieval \\
\(\text{Score}(\hat{o})\) & Re-ranking Function & Score for final personalization/trend re-ranking:  
\(\text{Score}(\hat{o}) = \alpha(u)\cdot \text{Preference}(\hat{o}) + \beta(u)\cdot \text{Trend}(\hat{o})\). & Preference-Trend Re-Ranking\\
\midrule  
\multicolumn{4}{l}{\textbf{Evaluation Metrics}} \\
\(\text{R@K}\) & Recall@K & \(\text{R@K} = \frac{1}{N} \sum_{i=1}^N \mathbb{I}_{\text{rel}}(i,\ \text{@K})\). Emphasizes how many samples have at least one relevant item in the top-K. & Evaluation Section \\
\(\text{MAP}\) & Mean Average Precision & \(\text{MAP} = \frac{1}{N} \sum_{i=1}^N \frac{1}{n_i} \sum_{k=1}^{L_i} \frac{\text{P}@k \cdot \mathbb{I}_{\text{rel}}(i,k)}{k}\). Reflects averaged precision over ranked positions. & Evaluation Section \\
\(\mathbb{I}_{\text{rel}}(\cdot)\) & Indicator Function & Has two forms: \(\mathbb{I}_{\text{rel}}(i,\ \text{@K})=1\) if any ground-truth item is in top-K; \(\mathbb{I}_{\text{rel}}(i,k)=1\) if the item at rank \(k\) is relevant. & Evaluation Section \\
\(n_i,\ L_i,\ N\) & Evaluation Constants & \(N\): total test samples; \(n_i\): number of ground-truth relevant items for sample \(i\); \(L_i\): recommendation list length for sample \(i\). & Evaluation Section \\
\bottomrule  
\end{tabular}
}
\end{table*}

\section{PAFA Knowledge Base}
In this chapter, we will further introduce the data sources, data cleaning processes, and core libraries involved in building the PAFA knowledge base. The relevant knowledge base and code will be open-sourced after the official publication.

\subsection{Data Collection and Processing}
 The PAFA knowledge base is built on multi-source heterogeneous data, covering structured and unstructured data acquisition, processing, and fusion. This section outlines data sources, preprocessing workflows, and standardization methods to ensure the knowledge base's professionalism, timeliness, and consistency.

The knowledge base's quality is supported by diverse and authoritative data sources, categorized as follows:
\emph{Academic Knowledge}. Incorporates over 120 specialized books and thesis. LLMs VLMs are utilized for semantic analysis, extracting entities, relationships, and attributes to build structured representations. Color matching references scientific models (e.g., analogous, contrasting colors), and silhouette knowledge follows classic design principles.

\emph{Industry Trends}. Automatically extracts fashion trends from reports by over 50 authoritative institutions (e.g., WGSN, Pantone), runway cases, and street photography. Dynamic trends are tracked via regular data crawling and parsing.

\emph{User Behavior Data}. Captures user outfit records history, and interaction preferences (e.g., favorites) from open-source datasets to build personalized style profiles, adding a dynamic, user-specific dimension. User preference modeling identifies scenario preferences, color tendencies, and silhouette choices while maintaining data privacy.

\subsubsection{Data Preprocessing and Standardization}
Data preprocessing involves cleaning, deduplication, and standardization to resolve inconsistencies in terminology and representations. For text data, large language models (LLMs) are used to extract structured attributes, perform zero-shot extraction of clothing attribute triplets, and leverage multi-turn reasoning to identify professional fashion terminology. LLMs also infer implicit fashion knowledge, such as unifying variations in expressions of the same concept (e.g., mapping "navy blue" and "dark blue" to a single color category) and constructing synonym networks to standardize terminology. For image data, multimodal models like GPT-4o are employed to extract structured attribute descriptions.

Cross-modal alignment between text and images is achieved using CLIP-based methods, where image and text representations are integrated into a unified semantic space by optimizing cosine similarity. Additionally, diffusion model-based techniques enable bidirectional conversion between text and image features, supporting tasks such as generating visual features from text descriptions and producing professional text descriptions for images. Adaptive temporal filtering ensures the timeliness of trend data by removing noise while preserving long-term patterns, further enhancing the multimodal completeness and consistency of the knowledge base.

\subsection{Core Sub-Knowledge Bases}

The PAFA knowledge base consists of six specialized sub-knowledge bases, each focused on core areas with defined modeling goals and data strategies. This modular structure boosts flexibility while advancing coverage, expertise, and personalization.

\subsubsection{Scenario-Style}
The Scenario-Style Repository establishes a systematic framework for quantifying the compatibility of clothing styles across different life contexts, providing contextualized knowledge support for intelligent recommendation systems.

Based on the analysis of major life scenarios and style dimensions, we have constructed a scenario-style matrix $\mathbf{M}_{s} \in \mathbb{R}^{m \times n}$, where $m=5$ primary scenarios (business, social, casual, sports, and special occasions) and $n=24$ detailed styles. Each element $m_{ij}$ in the matrix represents the compatibility value between scenario $i$ and style $j$, ranging from $[0,1]$, derived from normalized expert ratings. The core structure can be simplified as:


\begin{equation} 
\mathbf{M}_{scene} = \begin{bmatrix}
\text{Business} & \text{Minimalist} & \text{Classic} & \cdots \\
\text{Social} & \text{Romantic} & \text{Elegant} & \cdots \\
\text{Casual} & \text{Street} & \text{Sporty} & \cdots \\
\text{Special} & \text{Vintage} & \text{Avant-garde} & \cdots \\
\vdots & \vdots & \vdots & \ddots
\end{bmatrix}
\end{equation}

\subsubsection{Color Matching Principle}
The Color Matching Rule Repository constitutes a systematic framework of color pairing principles based on color theory, designed to provide precise color guidance for recommendation systems. This repository leverages theoretical foundations from CIE LAB color space, HSV color space, and the Munsell color system, quantifying harmony between colors through a harmony function $harmony(c_1, c_2) \rightarrow [0,1]$, where color differences are calculated using $\Delta E_{ab}^ = \sqrt{(L_1-L_2)^2 + (a_1-a_2)^2 + (b_1-b_2)^2}$.
This rule system encompasses: (1) Monochromatic principles, maintaining consistent hue while varying saturation and brightness to create harmonious variations; (2) Analogous principles, selecting Analogous colors on the color wheel to form natural transitions; (3) Complementary principles, choosing opposite colors on the color wheel, particularly triggering high visual impact contrasts when $\Delta E_{ab}^ > 30$; (4) Triadic principles, selecting three equidistant colors on the color wheel to create balanced yet dynamic combinations; (5) Neutral principles, incorporating colors with moderate luminance as foundational or transitional elements. This rule system also integrates color psychology factors, recognizing that warm tones (red, orange, yellow) typically convey enthusiasm and energy, while cool tones (blue, purple) communicate calmness and professionalism, thereby extending the application scope through an emotional dimension. Additionally, the repository incorporates a color harmony scoring mechanism that calculates the overall coherence of color schemes based on rule compliance, providing quantitative decision support for recommendation systems. Through this systematic color rule repository, recommendation systems can generate color suggestions that both adhere to color theory principles and satisfy diverse aesthetic requirements.

\subsubsection{Silhouette Matching Principle}

The Silhouette Matching Rule Repository serves as a crucial knowledge component in garment recommendation systems, providing a structured representation of fashion design expertise through formalized descriptions of the matching relationships between garment silhouettes and human body shapes. This repository comprises three fundamental elements.  

The study employs multidimensional feature vectors to characterize garment silhouettes, establishing a comprehensive classification system. The \emph{H-type silhouette} ($H_{silhouette}$) features a straight vertical contour with consistent width from top to bottom, while the \emph{X-type silhouette} ($X_{silhouette}$) exhibits a narrowed waist and symmetrical expansion at both upper and lower sections. The \emph{A-type silhouette} ($A_{silhouette}$) gradually widens downward, forming a triangular shape, whereas the \emph{Y-type silhouette} ($Y_{silhouette}$) has broad shoulders that taper downward, creating an inverted triangular contour. The \emph{O-type silhouette} ($O_{silhouette}$) presents a rounded overall curve without distinct waist definition. Each silhouette is represented by a vector $\boldsymbol{s} = (s_1, s_2, ..., s_n)$, where $s_i$ denotes key feature parameters such as shoulder width ratio, waist narrowing degree, and hem flare angle, serving as quantitative indicators.  

The balancing principles for coordinating upper and lower garment silhouettes are grounded in visual psychology's proportion harmony theory, formalized as:  
\begin{equation}     
balance(silhouette_{top}, silhouette_{bottom}) \rightarrow harmony_score
\end{equation}
These principles primarily adhere to the following principles:  
Contrast Principle: The $harmony_score$ increases when upper and lower garments create visual contrast, e.g., 
\begin{equation}     
balance(tight\_top, loose\_bottom) > balance(tight\_top, tight\_bottom)
\end{equation}
Visual Stability Principle: The $harmony_score$ improves when silhouette combinations maintain visual balance, e.g., 
\begin{equation} 
balance(Y_{silhouette}, A_{silhouette}) > balance(Y_{silhouette}, H_{silhouette})
\end{equation}
Golden Ratio Principle: The $harmony_score$ reaches its optimum when the upper-to-lower garment proportion approximates the golden ratio.

\subsubsection{Outfit Cases}

The Outfit Repository, serving as a core knowledge component of recommendation systems, employs a multimodal representation architecture to store and organize clothing coordination knowledge. Each outfit case $C_i$ is structured as a triplet $(I_i, T_i, M_i)$, representing visual outfit representation, structured semantic description, and metadata attribute set, respectively. The system ensures consistency between visual and semantic representations through cross-modal alignment mechanisms, satisfying the constraint condition $d(\Phi_{img}(I_i), \Phi_{text}(T_i)) < \epsilon$.

The repository characterizes clothing item features through a multi-layer attribute model, including key dimensions $A_{item} = \{a_{color},\allowbreak a_{pattern}, \allowbreak a_{fabric}, a_{silhouette}, a_{style}\}$. Meanwhile, a rule extraction module $\mathcal{R}$ derives coordination knowledge from annotated cases, formalized as rule sets encompassing color harmony, proportion balance, and style consistency. Each rule can be expressed as a conditional probability: $P(Compatible|a_i^{top}, a_j^{bottom}) = \allowbreak f_{\theta}(a_i^{top}, \allowbreak a_j^{bottom})$, where $f_{\theta}$ represents a parameterized compatibility evaluation function. At the implementation level, the repository employs vector database indexing to accelerate similar case retrieval.

\subsubsection{Fashion Trend}
The Trend Knowledge Base integrates data from authoritative fashion resources, such as WGSN reports, fashion week coverage, Pantone color releases, industry magazines, and social media trends. These data are cleansed and structured to extract key elements like popular styles, seasonal colors, and mainstream patterns, forming standardized trend knowledge entries. Its value lies in dynamically supporting recommendation systems through regular updates to reflect the latest market trends. This enables the Retrieval-Augmented Generation (RAG) system to provide precise recommendations that align with personal preferences and current trends, enhancing user experience and content relevance. Additionally, the knowledge base maps trend elements to product features, transforming abstract concepts into concrete recommendation logic.

\subsubsection{Personal Preference}
The personal preference repository identifies statistically significant aesthetic patterns and selection paradigms by mining group patterns from cross-user historical outfit data, thereby constructing a behavioral consensus-based style preference model. The system extracts common patterns across three dimensions: first, it analyzes style clustering characteristics in high-frequency scenarios (e.g., commuting scenarios predominantly feature low-saturation minimalist styles); second, it quantifies the distribution of group preferences for hue and brightness (revealing that workplace groups exhibit significantly higher selection probabilities for earth tones compared to casual scenarios); and third, it identifies pattern combination principles through clustering of tailoring data (e.g., H-line outerwear and tapered pants show strong correlation in business settings).

\end{document}